\documentstyle[twocolumn,aps,psfig]{revtex}

\begin{document}
\draft
\flushbottom
\twocolumn[
\hsize\textwidth\columnwidth\hsize\csname @twocolumnfalse\endcsname

\title{ Surface plasmon toy-models of black holes and wormholes.}
\author{Igor I. Smolyaninov, Christopher C. Davis}
\address{ Department of Electrical and Computer Engineering \\
University of Maryland, College Park,\\
MD 20742}
\date{\today}
\maketitle
\tightenlines
\widetext
\advance\leftskip by 57pt
\advance\rightskip by 57pt

\begin{abstract}
Surface plasmons at metal interfaces are collective excitations of the conduction electrons and the electromagnetic field. They exist in "curved
three-dimensional space-times" defined by the shape of the metal surface and the spatial distribution of the dielectric constant near the surface. Here we show that surface plasmon toy models of many non-trivial space-time metrics, such as wormholes and black holes, can be easily built and studied in experiments. For example, a droplet of dielectric on the metal surface behaves as a black hole for surface plasmons within a substantial frequency range. On the other hand, a nanohole in a thin metal membrane may be treated as a wormhole connecting two "flat" surface plasmon worlds located on the opposite interfaces of the membrane. 
\end{abstract}

]
\narrowtext

\tightenlines

The realization that solid-state toy models may help in an understanding of electromagnetic phenomena in curved space-time has led to considerable recent effort in developing toy models of electromagnetic \cite{1} and sonic \cite{2} black holes. In the case of media electrodynamics this is possible because of an analogy between the propagation of light in matter and in curved space-times: it is well known that Maxwell's equations in a general curved space-time background $g_{ik}(x,t)$ are equivalent to the phenomenological Maxwell equations in the presence of a matter background with nontrivial electric and magnetic permeability tensors $\epsilon _{ij}(x,t)$ and $\mu _{ij}(x,t)$ \cite{3}. In this analogy, the event horizon corresponds to a surface of singular electric and magnetic permeabilities, so that the speed of light goes to zero, and light is "frozen" near such a surface. In the absence of established quantum gravitation theory the toy models are helpful in understanding electromagnetic phenomena in curved space-times, such as Hawking radiation \cite{4} and the Unruh effect \cite{5}. Very recently it was also realized \cite{6,7} that many results obtained in the optics of random media may be applicable to the case of light propagation in stochastic space-time metrics. Thus, an exchange of ideas between media electrodynamics and gravitation theory proves to be useful for both fields. Unfortunately, up to now all the suggested electromagnetic black hole toy-models were very difficult to realize and study experimentally, so virtually no experimental work has been done in this field. For example, the electromagnetic toy black hole suggested in \cite{1} requires a peculiar spatial distribution of temperature inside a ferroelectric and ferromagnetic crystal, which would be very difficult to create and maintain over any reasonable period of time. 

In this Letter we propose novel electromagnetic toy models of wormholes and black holes, which are based on surface plasmons and are very easy to build and study. We are going to consider a close analogy between the optics of surface plasmons propagating along metal interfaces, which are curved and/or covered with dielectric films, and the field theory in a curved space-time background. Surface plasmons are collective excitations of the conductive electrons and the electromagnetic field \cite{8}. They exist in "curved three-dimensional space-times" defined by the shape of the metal-dielectric interface. Since in many experimental geometries surface plasmons are weakly coupled to the outside world (to free-space photons, phonons, single-electron excitations, etc.) it is reasonable to treat the physics of surface plasmons separately from the rest of the surface and bulk excitations, so that a field-theory of surface plasmons in a curved space-time background may be considered. For example, a nanohole in a thin metal membrane may be treated as a "wormhole" connecting two "flat" surface plasmon worlds located on the opposite interfaces of the membrane \cite{9}. This nanohole may be filled with a material with some nontrivial electric and magnetic permeability tensors $\epsilon _{ij}(x,t)$ and $\mu _{ij}(x,t)$, so that nontrivial space-time metrics can be emulated. This point of view allowed us to justify and explore an analogy between the nonlinear optics of cylindrical surface plasmons of nanowires and nanoholes and the lower-dimension Kaluza-Klein theories \cite{9}. 

On the other hand, near the plasmon resonance (which is defined by the condition that $\epsilon _m(\omega )=-\epsilon _d$ at the metal-dielectric interface, where $\epsilon _m(\omega )$ and $\epsilon _d$ are the dielectric constants of metal and dielectric, respectively \cite{8}) the surface plasmon velocity vanishes, so that the surface plasmon "stops" on the metal surface, and the surface charge and the normal component of the electric field diverge. For a given frequency of light, the spatial boundary of the plasmon resonance ("the event horizon" of our toy model) may be defined at will using the geometry $\epsilon _d(x,y)$ of the absorbed layer of dielectric on the metal surface. Thus, the plasmon resonance becomes a natural candidate to emulate the event horizon of a black hole. As a result, toy two-dimensional surface plasmon black holes can be easily produced and studied. In what follows we are going to show that consideration of such surface plasmon toy wormholes and black holes may be useful both for the fields of nanooptics and for gravitation theory. 

Let us consider in detail the dispersion law of a surface plasmon (SP), which propagates along the metal-dielectric interface. The SP field decays exponentially both inside the metal and the dielectric. Inside the dielectric the decay exponent is roughly equal to the SP wave vector. As a first step let us assume that both metal and dielectric completely fill the respective $z<0$ and $z>0$ half-spaces. In such a case the dispersion law can be written as \cite{8} 

\begin{equation}  
k^2=\frac{\omega ^2}{c^2}\frac{\epsilon _d\epsilon _m(\omega )}{\epsilon _d+\epsilon _m(\omega)} ,
\end{equation}

where we will assume that $\epsilon _m=1-\omega _p^2/\omega ^2$ according to the Drude model, and $\omega _p$ is the plasma frequency of the metal. This dispersion law is shown in Fig.1(b) for the cases of metal-vacuum and metal-dielectric interfaces. It starts as a "light line" in the respective dielectric at low frequencies and approaches asymptotically $\omega =\omega _p/(1+\epsilon _d)^{1/2}$ at very large wave vectors. The latter frequency corresponds to the so-called surface plasmon resonance. Under the surface plasmon resonance conditions both phase and group velocity of the SPs is zero, and the surface charge and the normal component of the electric field diverge. Since at every wavevector the SP dispersion law is located to the right of the "light line", the SPs of the plane metal-dielectric interface are decoupled from the free-space photons due to the momentum conservation law.   

If a droplet of dielectric (Fig.1(a)) is placed on the metal surface, the SP dispersion law will be a function of the local thickness of the droplet. Deep inside the droplet far from its edges the SP dispersion law will look similar to the case of a metal-dielectric interface, whereas near the edges (where the dielectric is thinner) it will approach the SP dispersion law for the metal-vacuum interface. As a result, for every frequency between $\omega _p/(1+\epsilon _d)^{1/2}$ and $\omega _p/2^{1/2}$ there will be a closed linear boundary inside the droplet for which the surface plasmon resonance conditions are satisfied. Let us show that such a droplet of dielectric on the metal interface behaves as a "surface plasmon black hole" in the frequency range between $\omega _p/(1+\epsilon _d)^{1/2}$ and $\omega _p/2^{1/2}$, and that the described boundary of the surface plasmon resonance behaves as an "event horizon" of such a black hole. 

Let us consider a SP within this frequency range, which is trapped near its respective "event horizon", and which is trying to leave a large droplet of dielectric (see Fig.2). The fact that the droplet is large means that ray optics may be used. Since the component of the SP momentum parallel to the droplet boundary has to be conserved, such a SP will be totally internally reflected by the surface plasmon resonance boundary back inside the droplet at any non-zero angle of incidence. This is a simple consequence of the fact that near the "event horizon" the effective refractive index of the droplet for surface plasmons is infinite (according to eq.(1), both phase and group velocity of surface plasmons is zero at surface plasmon resonance). On the other hand, even if the angle of incidence is zero, it will take the SP infinite time to leave the resonance boundary. Thus, the droplet behaves as a black hole for surface plasmons, and the line near the droplet boundary where the surface plasmon resonance conditions are satisfied plays the role of the event horizon for surface plasmons. 

The toy black holes described above are extremely easy to make and observe. In our experiments a small droplet of glycerin was placed on the gold film surface and further smeared over the surface using lens paper, so that a large number of glycerin microdroplets were formed on the surface (Fig.3(a)). These microdroplets were illuminated with white light through the glass prism (Fig.1(a)) in the so-called Kretschman geometry \cite{8}. The Kretschman geometry allows for efficient SP excitation on the gold-vacuum interface due to phase matching between the SPs and photons in the glass prism. As a result, SPs were launched into the gold film area around the droplet. Photograph taken under a microscope of one of such microdroplets is shown in Fig.3(b). The white rim of light near the edge of the droplet is clearly seen. It corresponds to the effective SP event horizon described above. Near this toy event horizon SPs are stopped or reflected back inside the droplet. In addition, a small portion of the SP field may be scattered out of the two-dimensional surface plasmon world into normal three-dimensional photons. These photons produced the image in Fig.3(b). If we are to continue our analogy with real black holes in the worlds with other than three spatial dimensions, we should state that near the event horizon the surface plasmons are scattered from their two-dimensional world into "extra dimensions" (which in this case is our normal three-dimensional space).      

Unfortunately, the described toy SP black hole model does not work outside the frequency range between $\omega _p/(1+\epsilon _d)^{1/2}$ and $\omega _p/2^{1/2}$. On the other hand, this is a common feature of every electromagnetic toy black hole model suggested so far. All such toy models necessarily work only within a limited frequency range. However, the ease of making and observing such toy SP black holes makes them a very promising research object. If we forget about the language of "black holes" and "event horizons" for a moment, the SP optics phenomenon represented in Fig.3(b) remains a potentially very interesting effect in surface plasmon optics. Namely, this photo shows the existence of a two-dimensional SP analog of whispering gallery modes, which are well-known in the optics of light in droplets and other spherical dielectric particles. Whispering gallery modes in liquid microdroplets are known to substantially enhance nonlinear optical phenomena due to cavity quantum electrodynamic effects \cite{10}. One may expect even higher enhancement of nonlinear optical mixing in liquid droplets on the metal surfaces due to enhancement of surface electromagnetic field inherent to surface plasmon excitation, and in addition, due to accumulation of SP energy near the surface plasmon event horizons at the droplet boundaries. This strong enhancement of nonlinear optical effects in liquid droplets may be very useful in chemical and biological sensing applications. 

Looking back at the consequences of our model for gravitation theory, we may anticipate that such effects as recently predicted optical second harmonic generation near the black hole event horizon \cite{7} can be easily observed in experiments with toy SP black holes. One can even look forward to modeling of more exotic (and maybe more realistic) gravitation theory situations where a wormhole exists inside a black hole. In the more prosaic language of surface plasmon optics this situation corresponds to nanoholes drilled in a free standing metal membrane and covered with droplets of dielectric (Fig.4(a,c)). In fact, our recent experiments on single-photon tunneling \cite{11} and optical control of photon tunneling \cite{12} were performed with exactly these kinds of samples. Theoretical analysis of these experiments performed in \cite{9} indicated some usefulness of the gravitational theory analogy: parallels can be drawn between the nonlinear optics of surface plasmons in cylindrical nanoholes (cylindrical surface plasmons) and areas of gravitation theory such as Kaluza-Klein theories \cite{9}. 
   
This analogy stems from the way in which electric charges are introduced in the original five-dimensional Kaluza-Klein theory (see for example \cite{13,14} and Fig.4(b)). In this theory the electric charges are introduced as chiral (nonzero angular momentum) modes of a massless quantum field, which is quantized over the cyclic compactified fifth dimension. Electromagnetic forces between the electric charges appear as nonlinear coupling of these chiral modes, so that four-dimensional electrodynamics described by the Maxwell equations may be understood as nonlinear optics of these modes. Similar Kaluza-Klein theories may be formulated in lower-dimensional space-times. Such theories reproduce electrodynamics of electric charges in worlds with less than three spatial dimensions. Since nonlinear interactions of surface plasmons of a cylindrical nanowire or a nanohole look as if they occur in a space-time which besides an extended z-coordinate has a small "compactified" angular $\phi $-dimension along the circumference of the cylinder, the theory of cylindrical surface plasmon (CSP) mode propagation and interaction appears to be similar to the three-dimensional Kaluza-Klein theory: solutions of the nonlinear Maxwell equations for interacting cylindrical surface plasmons may be found \cite{9}, which behave as interacting Kaluza-Klein charges. According to these solutions, higher $(n>0)$ CSP modes posses quantized effective chiral charges proportional to their angular momenta $n$. In a metal nanowire these slow moving effective charges exhibit long-range interaction via exchange of fast massless CSPs with zero angular momentum. These zero angular momentum CSPs may be considered as massless quanta of the gyration field (the field of the gyration vector $\vec{g}$), which relates the $\vec{D}$ and $\vec{E}$ fields in an optically active medium \cite{15}: 

\begin{equation}  
\vec{D}=\epsilon \vec{E}+i\vec{E}\times \vec{g} ,
\end{equation}

In order for such a physical picture to be valid the medium in or around the nanohole or nanowire should be chiral or optically active, and exhibit the magneto-optical effect. This condition is fulfilled automatically in the case of any metal \cite{15}, since all metals are optically active in the presence of a magnetic field. Thus, optical experiments performed on nanowires and nanoholes may become the proving ground for experimental testing of many ideas in theoretical physics, such as compactified extra dimensions and wormholes.

Detailed illustration of how the nonlinear optics of cylindrical surface plasmons may be formulated in a way that is similar to Kaluza-Klein theories can be found in \cite{9}. The value of the Kaluza-Klein and black hole models for the field of nanooptics is quite apparent: these models provide very useful intuitive guides for finding solutions of nonlinear Maxwell equations in the situations when nonlinear interactions are strong and can not be considered as small perturbations. In such situations, often one can only guess the general form of the solutions, based on the comparison of the nonlinear system of interest with other better understood nonlinear systems described by similar equations. We can also ask if experimental and theoretical studies of nanoholes and microdroplets can be useful for gravitation theory. Some arguments can be made that this might be the case. For example, the solutions for the electromagnetic eigenmodes of the wormholes and their interaction are often sought for very narrow wormholes, which are understood as strings connected to D-branes (D-dimensional membranes) \cite{16}. The multi-component Kaluza-Klein charges of the wormholes in such configurations have some resemblance to the multi-component chiral charges of metal nanowires and nanoholes described in \cite{9}. Thus, solid state models may provide some useful insights. It is customary in theoretical description of wormholes in higher-dimensional Kaluza-Klein theories to distinguish between electric and magnetic Kaluza-Klein charges and fields \cite{17}. According to this classification, we should identify the chiral charges of the cylindrical surface plasmons as magnetic charges, while (quite naturally) the conductive electrons in the metal nanowire perform the role of the electric Kaluza-Klein charges \cite{9}. Another important question, which follows from the analogy between the nanoholes and the wormholes arises from the fact that many optical waveguides (from the electrodynamics point of view a wormhole is a waveguide) have cut-off frequencies for the propagating electromagnetic modes. For example, unlike metal nanowires, the nanoholes do not support propagating electromagnetic modes with small frequencies. This may mean that in gravitation theory an attempt to describe a wormhole using a frequency-independent space-time metric may fail in some cases. Excitations with small energies below the wormhole cut-off may not see the wormhole throat at all.      

In conclusion, we have introduced and observed experimentally surface plasmon analogues of such nontrivial space-time topologies as black holes and wormholes. We anticipate that further consideration of such analogues will produce a mutually beneficial exchange of ideas between nanooptics and gravitation theory.

Figure captions.

Fig.1 (a) Experimental geometry of surface plasmon toy black hole observation. (b) Surface plasmon dispersion law for the cases of metal-vacuum and metal-dielectric interfaces

Fig.2 A surface plasmon trapped inside a droplet near the effective "event horizon": The projection of surface plasmon momentum parallel to the droplet edge must be conserved. Due to effectively infinite refractive index near the droplet edge surface plasmons experience total internal reflection at any angle of incidence. 

Fig.3 Microscopic photographs of a toy surface plasmon black hole. (a) Droplet of glycerin on a gold film surface (illuminated from the top). The droplet diameter is approximately 15 micrometers. (b) The same droplet illuminated in the Kretschman geometry, which provides efficient coupling of light to surface plasmons on the gold-vacuum interface (Fig.1(a)). The white rim around the droplet boundary corresponds to the effective surface plasmon "event horizon". 

Fig.4 (a) A nanohole in a metal membrane may be treated as a wormhole for surface plasmons which exist on the flat top and bottom interfaces of the membrane. Alternatively, this figure may represent a nanowire connecting two metal-vacuum interfaces. (b) Similarity between Kaluza-Klein electric charges and cylindrical surface plasmons. (c) Array of 20 nm diameter nanoholes drilled in a free standing gold membrane, which has been studied experimentally \cite{12}. Such nanoholes may be treated as surface plasmon wormholes.

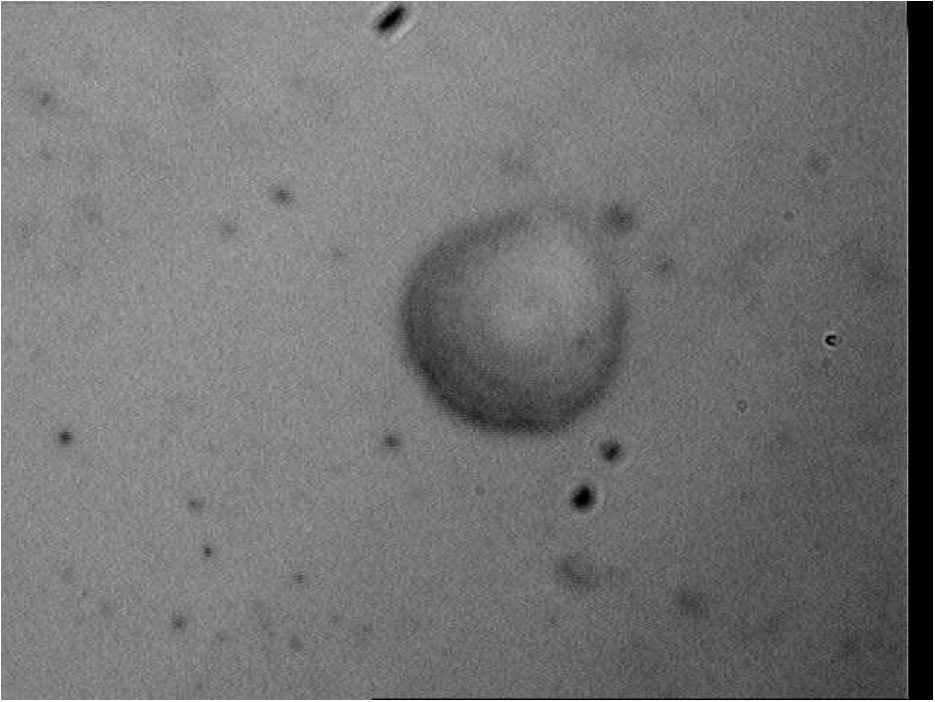
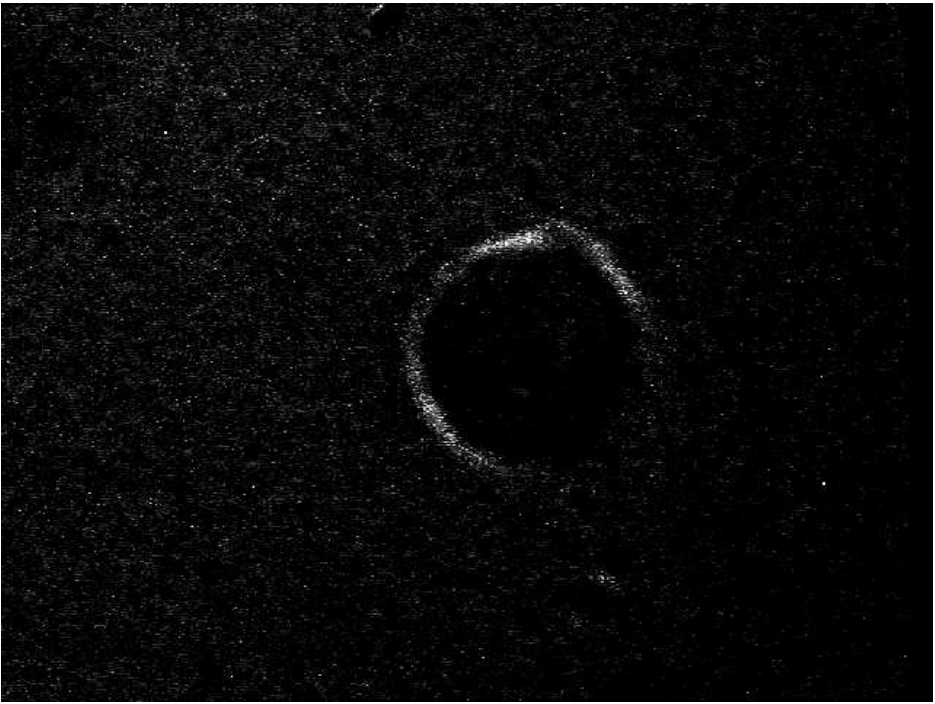
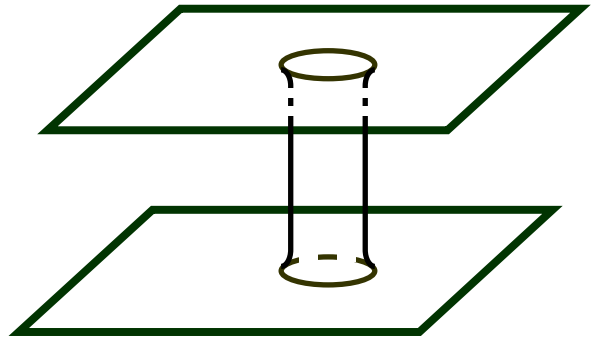
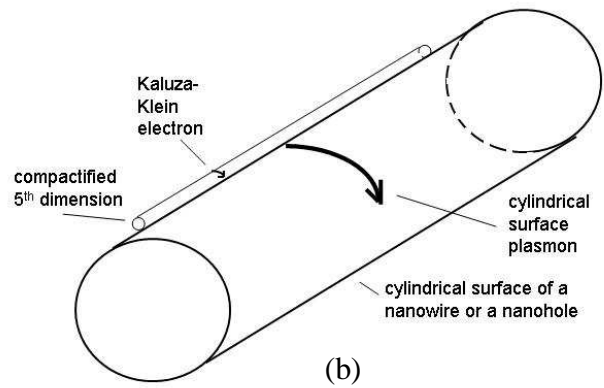
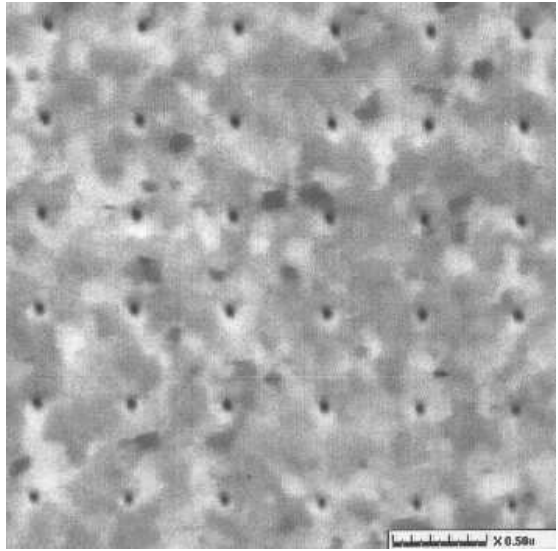
\end{document}